# Structure and energetics of solvated ferrous and ferric ions: Car-Parrinello molecular dynamics in the DFT+U formalism

P. H.-L. Sit[1], M. Cococcioni[2] and Nicola Marzari[2]

[1]*Department of Physics, Massachusetts Institute of Technology, Cambridge, MA 02139, USA,*

[2]*Department of Materials Science and Engineering, Massachusetts Institute of Technology, Cambridge, MA 02139, USA*

(Dated: January 2, 2007)

## Abstract

We implemented a rotationally-invariant Hubbard U extension to density-functional theory in the Car-Parrinello molecular dynamics framework, with the goal of bringing the accuracy of the DFT+U approach to finite-temperature simulations, especially for liquids or solids containing transition-metal ions. First, we studied the effects on the Hubbard U on the static equilibrium structure of the hexa-aqua ferrous and ferric ions, and the inner-sphere reorganization energy for the electron-transfer reaction between aqueous ferrous and ferric ions. It is found that the reorganization energy is increased, mostly as a result of the Fe-O distance elongation in the hexa-aqua ferrous ion. Second, we performed a first-principles molecular dynamics study of the solvation structure of the two aqueous ferrous and ferric ions. The Hubbard term is found to change the Fe-O radial distribution function for the ferrous ion, while having a negligible effect on the aqueous ferric ion. Moreover, the frequencies of vibrations between Fe and oxygen atoms in the first-solvation shell are shown to be unaffected by the Hubbard corrections for both ferrous and ferric ions.

## I. INTRODUCTION

The combination of density-functional theory (DFT) and molecular dynamics, as embodied in the milestone paper of Car and Parrinello [1], has introduced a new era in the accurate and predictive simulation of material properties. This is especially relevant in the treatment of liquids, where finite-temperature molecular dynamics simulations are often needed to extract observables and ensemble averages. Car-Parrinello simulations are most commonly implemented in a pseudopotential, plane-wave formalism [2], to take advantage of the uniform resolution of plane waves across space, the absence of Pulay contributions to the Hellmann-Feynman forces, and the reliability of current pseudopotentials, especially in the ultrasoft formulation [3], that remove the need for an explicit treatment of the core electrons. Local or semi-local approximations to the exchange correlation functional, such as the widely-employed PBE [4] generalized-gradient approximation, are also directly amenable to a Lagrangian formulation, where the functional derivatives of the exchange correlation functional can be inexpensively calculated in a plane-wave basis. While very successful, such approximations are known to display qualitative or quantitative failures in some special cases. The most notable failure is the presence of self-interaction [5, 6]: at variance with e.g. Hartree-Fock, the exchange-correlation functional does not exactly cancel out the contributions of each one-electron orbital interacting with its own charge-density distribution. Broadly related to this failure is the inability to treat strongly-correlated materials (usually systems with partially-filled $d$ or $f$-valence shells): in reality, the $d$ or $f$-electrons feel a strong Coulomb repulsions in the presence of other localized electrons, and the electron motion is known to be "correlated" due to this Coulomb interaction. In these cases, the tendency embodied in local or semi-local exchange-correlation functionals to de-localize and over-hybridize enhances the metallic character and band picture of these oxides, leading to notable failures, from magnetic ordering to metallic-versus-insulating behavior [7]. A Hubbard U extension to DFT (DFT+U) [8–10] has been proposed to better describe strong electronic correlations - where the goal of capturing electron-electron interactions is aided by a Hubbard U term, which models the on-site repulsion. This scheme has been shown to be a simple and effective way to describe strongly-correlated systems from first-principles [10–13]. In addition, it was recently shown that U is not an empirical fitting parameter, but a truly ab-initio quantity that can be derived from the bare and screened linear response of the system under consideration to a change in occupations [14]. Recently, we have argued that such an approach can also be very successful in transition-metal complexes, where the U term leads to an improved description of the energies and structure of compounds with single-site or few-site localized electrons, thanks to an improved description of self- or intra-atomic interactions [15].

A Car-Parrinello implementation of a DFT+U functional would naturally lead to an improved description of the finite-temperature properties of strongly-correlated oxides, or of transition metal complexes in strong ligand fields. In this paper, we briefly summarize and review the fundamentals of such approaches, and we present our implementation of a generalized DFT+U Car-Parrinello Lagrangian, that we implemented in the public QUANTUM-ESPRESSO package [16]. This approach was extensively tested, showing an excellent conservation of the constant of motion, and was applied to the study of ferrous and ferric aqua ions.

In the next section, we briefly summarize the simplified, rotational invariant DFT+U scheme developed by Cococcioni *et al.* [10]. In section III, we discuss the calculation of the Hubbard U (which describes the strength of the correlation) directly from first-principles using linear-response [14]. This rotational invariant DFT+U scheme has recently been shown to accurately describe a wide range of transition metal systems [10, 17, 18], and especially transition-metal complexes [15]. In section IV, we study the effects of the Hubbard U correction on the solvation structure of ferrous and ferric ions in water and on the reorganization energy of aqueous ferrous-ferric self-exchange reactions. We then discuss in section V the implementation of DFT+U in a Car-Parrinello molecular dynamics framework. Finally, in section VI we investigate the effects of U on the solvation structure of ferrous and ferric ions in water.

## II. A SIMPLIFIED ROTATIONALLY-INVARIANT SCHEME

In order to treat the effects of correlation, a correction functional $E^U[\{n_{mm'}^{I\sigma}\}]$ is added to the standard DFT functional $E^{DFT}[n]$ [8, 9]:

$$\begin{aligned} E^{DFT+U}[n] &= E^{DFT}[n] + E^U[\{n_{mm'}^{I\sigma}\}] \\ &= E^{DFT}[n] + E^{Hub}[\{n_{mm'}^{I\sigma}\}] - E^{dc}[\{n^{I\sigma}\}], \end{aligned} \qquad (1)$$

where $n$ is the total electronic density, $n^{I\sigma}_{mm'}$ is a localized-orbitals occupation matrix and $n^{I\sigma} = \sum_m n^{I\sigma}_{mm}$ is the trace of the occupation matrix. In this approach, a small number of localized orbitals is selected and the electronic correlation associated to them is treated in a special way. These degrees of freedom are described by the occupation matrix

$$n^{I\sigma}_{mm'} = \sum_i f^\sigma_i \langle \psi^\sigma_i | S(\{R_I\}) | \phi^I_{m'} \rangle \langle \phi^I_m | S(\{R_I\}) | \psi^\sigma_i \rangle, \qquad (2)$$

where the $\{\psi^\sigma_i\}$ are the Kohn-Sham wavefunctions corresponding to state $i$ with spin $\sigma$, and $f^\sigma_i$ is the corresponding occupation number. The $\{\phi^I_m\}$ are localized orbitals selected; usually the valence atomic orbitals with angular momentum component $|lm_l\rangle$ of the atom sitting at site $I$ (the same wave functions are used for both spins). $E^{Hub}[\{n^{I\sigma}_{mm'}\}]$ then corresponds to on-site interactions between the correlated electrons, and $E^{dc}[\{n^{I\sigma}\}]$ is subtracted to avoid double-counting of the interactions contained both in $E^{Hub}[\{n^{I\sigma}_{mm'}\}]$ and, in an average way, in the local or semi-local approximations to DFT commonly used.

In the case of iron, the valence orbitals are the localized $d$-orbitals of iron atom and $l=2$. Since we will be using ultrasoft pseudopotentials to describe valence-core interactions, the augmentation charge operator $S(\{R_I\})$ is needed to describe in Eq. 2 the charge augmentation [3]. In $E^{dc}[\{n^{I\sigma}\}]$, we used the total, spin-polarized, occupation of the localized manifold $(n^{I\sigma})$. $E^U[\{n^{I\sigma}_{mm'}\}]$ can then be written as [10]

$$E^U[\{n^{I\sigma}_{mm'}\}] = \frac{U}{2} \sum_{I,\sigma} Tr[n^{I\sigma}(1-n^{I\sigma})]. \qquad (3)$$

As seen from Eq. 3, the Hubbard U functional introduces a penalty, tuned by the value of the U parameter, for partial occupation of the localized orbitals, thus favoring fully occupied or empty orbitals. Still, the interaction parameter U used in the calculations can be determined from first-principles, and not treated as an empirical fitting parameter.

### III. CALCULATION OF U FROM THE LINEAR RESPONSE APPROACH

We calculate the Hubbard U following the scheme proposed in Ref. [14]; we summarize the relevant steps below. The Hubbard U functional corrects and removes the spurious curvature due to the incomplete cancellation of the self-interactions in the DFT total energy with respect to the number of localized electrons. The second derivative of the total energy $\left(\frac{\partial^2 E^{DFT}}{\partial (n^I)^2}\right)$ corresponds to the effective curvature of the DFT total energy (here, $n^I$ is the total occupation of the localized orbitals, which can be calculated from the trace of the occupation matrix and summed over spins, $n^I = n^{I\uparrow} + n^{I\downarrow}$). However, this second derivative does not equal the Hubbard U. In fact, if we were to perturb a non-interacting system, we would still obtain a quadratic behavior of the total energy as a function of occupations. This is because the change in occupations changes the kinetic energy of the system, but this change should not be included because it does not originate from electron-electron interactions. Therefore, we use

$$U = \frac{d^2 E^{DFT}}{d(n^I)^2} - \frac{d^2 E_0^{DFT}}{d(n^I)^2}, \quad (4)$$

where the second derivative of $E_0^{DFT}$ removes the above-mentioned independent-electrons contribution (or bare response) from the unphysical curvature of the DFT functional that we want to correct. Since in actual calculations it is not convenient to constrain the total occupation of the localized orbitals, we force the occupations to vary by adding to the DFT functional a localized perturbation

$$E^{DFT}[\{\alpha\}] = \min_{n(r)}\{E^{DFT}[n^I] + \alpha n^I\}. \quad (5)$$

This new functional is $\alpha$ dependent; we can switch to an occupation-dependent energy functional by the Legendre transform

$$E^{DFT}[\{n^I\}] = \min_{\alpha}\{E^{DFT}[\alpha] - \alpha n^I\}. \quad (6)$$

From this, the double derivative of the total energy with respect to the occupation can be easily obtained as

$$\frac{d^2 E^{DFT}[n^I]}{d(n^I)^2} = -\frac{d\alpha[n^I]}{dn^I}. \quad (7)$$

In actual calculations, the quantity accessible is the response function $\chi = \frac{dn^I}{d\alpha}$ and the Hubbard U value equals

$$U = -\frac{d\alpha[n^I]}{dn^I} + \frac{d\alpha[n^I]}{dn_0^I} = (\chi_0^{-1} - \chi^{-1}), \quad (8)$$

where $\chi_0$ and $\chi$ are the bare and total response functions, respectively.

## IV. EFFECTS OF THE HUBBARD U TERM ON THE REORGANIZATION ENERGY OF ELECTRON TRANSFER

According to the Marcus theory of electron transfer [19, 20], only the reorganization energy $\lambda$ and the free energy of reaction are needed to fully characterize the diabatic free-energy surfaces if the solvent is in the linear response regime (i.e. the free-energy curves are parabolic). The reorganization energy $\lambda$ is the free energy cost to change from the equilibrium configurations of the product into the equilibrium configurations of the reactant without electron transfer having taken place. Previous studies [21, 22] have shown that the linear approximation holds in the case of ferrous-ferric self-exchange in water. Since the free energy of this symmetric reaction is zero, $\lambda$ is all what is needed to describe the free-energy surfaces.

In the case of ferrous and ferric ions solvated in water, the Hubbard U correction, being localized, is not going to be affected by water molecules outside the first solvation shell. For the same reason, this correction will have a small effect on the physical properties beyond the first solvation shell. Therefore, to study the effects of Hubbard U on the reorganization energy of electron transfer between ferrous and ferric ions, only the first solvation shells were considered. These have six water molecules in both cases, as shown by experiments [23, 24]. Using the scheme proposed in the previous section, we find U=5.6 eV for both the hexa-aqua ferrous and ferric ions.

With these value of U's, we obtain the inner-sphere contribution to the reorganization energy ($\lambda_{is}$) using the "4-point" method [25]. In this approach, the ionic geometry of a hexa-aqua ferrous cluster is first optimized in vacuum. One electron is then removed from the system and the total energy of the ferric ion in the optimized geometry of the ferrous ion cluster is calculated. A similar procedure is performed with the hexa-aqua ferric cluster and one extra electron is added to the system after ionic geometry optimization. $\lambda_{is}$ is then obtained as

$$\lambda_{is} = E^{Fe^{3+}}[Fe^{2+}] + E^{Fe^{2+}}[Fe^{3+}] - E^{Fe^{2+}}[Fe^{2+}] - E^{Fe^{3+}}[Fe^{3+}] \qquad (9)$$

where $E^{Fe^{2+}}[Fe^{2+}]$ and $E^{Fe^{2+}}[Fe^{3+}]$ are the total energies of a ferrous or ferric ion in the optimized geometry of the ferrous ion cluster. Similar notation is used for $E^{Fe^{3+}}[Fe^{2+}]$ and $E^{Fe^{3+}}[Fe^{3+}]$. We note in passing that in plane-wave codes employing periodic-boundary conditions comparing energies between systems with different charges requires extra care due to the finite Coulomb interactions between periodic images [26]. However, in the above procedure, these Coulomb interactions cancel out when calculating differences in energies.

We performed these calculations using PWscf [16] and ultrasoft pseudopotentials for iron, oxygen and hydrogen atoms at the PBE-GGA level. The ultrasoft pseudopotentials for O and H were taken from the standard distribution (H.pbe-rrkjus.UPF and O.pbe-rrkjus.UPF). We generated a 16-electron iron ultrasoft pseudopotential (Fe.pbe-sp-van_mit.UPF) using the Vanderbilt code [27], and a 2.5+ oxidation state. Plane-wave cutoffs for wavefunction and charge density are 25 and 200 Ryd, respectively. The ion cluster is in a unit cell of side length 15.9 Å.

TABLE I: The $d_{Fe-O}$'s of the relaxed structures for hexa-aqua ferrous and ferric ions with U=0 or 5.6 eV.

|  | U=0 eV | U=5.6 eV |
|---|---|---|
| Fe(H$_2$O)$_6^{2+}$ | $d_{Fe-O}$=2.11 Å  2.11 Å  2.14 Å  2.14 Å  2.14 Å  2.14 Å | $d_{Fe-O}$=2.15 Å  2.15 Å  2.18 Å  2.18 Å  2.18 Å  2.18 Å |
| Fe(H$_2$O)$_6^{3+}$ | $d_{Fe-O}$=2.05 Å  2.05 Å  2.05 Å  2.05 Å  2.05 Å  2.05 Å | $d_{Fe-O}$=2.06 Å  2.06 Å  2.06 Å  2.06 Å  2.06 Å  2.06 Å |

Using the above-described procedure, the inner sphere contribution $\lambda_{is}$ equals 0.50 eV when U=0 eV. With U=5.6 eV, we obtain $\lambda_{is}$=0.75 eV. This difference of 0.25 eV can be directly translated in the total reorganization energy since the contributions due to the outer water molecules are expected to be unaffected by the U correction. For the case when two ions are at 5.5 Å apart, the total reorganization free energy is 1.93 eV, as calculated from first-principles [22]. Therefore, after the Hubbard U correction (U=5.6 eV), $\lambda$ increases to 2.18 eV, which is in excellent agreement with the experimental value of 2.1 eV [25].

The change in the inner sphere reorganization energy can be related to the changes in relaxed structures of the clusters with a finite U. Table I shows the six Fe-O distances for hexa-aqua ferrous and ferric ions with U=0 and 5.6 eV. While the optimum structure for Fe(H$_2$O)$_6^{2+}$ is altered significantly with U=5.6 eV, there are minimal changes in the case of Fe(H$_2$O)$_6^{3+}$. The larger elongations for $d_{Fe-O}$ in the former case imply that a larger reorganization upon change in oxidation state of the ions, and thus a larger reorganization energy. The elongation of $d_{Fe-O}$ can be explained by the fact that the Hubbard U functional penalizes fractional occupations, thus weakening the hybridization between the lone-pair orbitals of water and the empty d-orbitals.

## V. IMPLEMENTATION OF DFT+U IN CAR-PARRINELLO MOLECULAR DYNAMICS

Due to the successes in the Hubbard U functional in studying systems with transition-metal ions [10, 14, 15], it is of great interest to be able to perform finite-temperature studies. In view of this, we implemented the Hubbard U formalism in the ultrasoft Car-Parrinello molecular dynamics [2] part of the Quantum-Espresso package [16]. We have also performed molecular dynamics simulations of ferrous and ferric ions solvated in water with the Hubbard U functional and we will present the results below.

By including a Hubbard U correction, the extended Car-Parrinello Lagrangian can be written as

$$\mathcal{L}_{CP} = \mu \sum_i f_i \int d\mathbf{r} \left| \dot{\Psi}_i(\mathbf{r}) \right|^2 + \frac{1}{2} \sum_I M_I \dot{\mathbf{R}}_I^2 - E^{DFT}[\{\Psi_i\}, \{\mathbf{R}_I\}]$$
$$- E^U[n_{m,m'}^{I,\sigma}] + \sum_{ij} \Lambda_{ij} \left( \int d\mathbf{r} \Psi_i^*(\mathbf{r}) \Psi_j(\mathbf{r}) - \delta_{ij} \right). \quad (10)$$

With this energy functional, wavefunctions and ions evolve according to the equations of motion,

$$M_I \ddot{R}_I = -\frac{\partial E^{DFT}[\{\psi_i\}, \{R_I\}]}{\partial R_I} - \frac{\partial E^U[n_{m,m'}^{I,\sigma}]}{\partial R_I} \quad (11)$$

$$\mu_i \ddot{\psi}_i^\sigma = -\frac{\delta E^{DFT}[\{\psi_i\}, \{R_I\}]}{\delta \psi_i^{\sigma*}} - \frac{\delta E^U[n_{m,m'}^{I,\sigma}]}{\delta \psi_i^{\sigma*}} + \sum_j \Lambda_{ij} \psi_j^\sigma. \quad (12)$$

Alternatively, in a Born-Oppenheimer dynamics, the Hellmann-Feynman theorem would apply, and forces on ions and wavefunctions could be written as

$$F^I = -\sum_{i,\sigma} \langle \psi_i^\sigma | \frac{d(\hat{H}^{DFT} + \hat{H}_i^U)}{dR_I} | \psi_i^\sigma \rangle \quad (13)$$

$$F^{\psi_i^\sigma} = -(\hat{H}^{DFT} + \hat{H}_i^U)\psi_i^\sigma + \sum_j \Lambda_{ij} \psi_j^\sigma. \quad (14)$$

In either case, the potential $\hat{H}_i^U$ due to the Hubbard functional is

$$\hat{H}_i^U = \frac{U}{2} \sum_{I,m,m'} f_i S(\{R_I\}) |\phi_m^I\rangle [\delta_{m,m'} - 2n_{m,m'}^{I,\sigma} \langle \phi_{m'}^I | S(\{R_I\})]. \quad (15)$$

# VI. DFT+U CAR-PARRINELLO MOLECULAR DYNAMICS SIMULATIONS OF AQUEOUS FERROUS AND FERRIC IONS WITH HUBBARD U CORRECTIONS

We performed molecular dynamics simulations with the Hubbard U corrections both to test the code and to study the effects of this functional on the dynamics. Detailed *ab-initio* molecular dynamics studies of aqua ferrous and ferric ions have been performed recently [28, 29] but the effects of Hubbard U corrections on the solvation structure and dynamics have not been investigated. We carried out two separate simulations with either ferrous or ferric ions solvated in 31 heavy water molecules. In Car-Parrinello molecular dynamics simulations, the constant of motion is the quantity to monitor in order to determine the accuracy of numerical implementations. The upper panel of Fig. 1 shows the constant of motion and the potential energy as a function of time for a molecular dynamics production run with a aqueous ferrous ion and U=5.6 eV. In these simulations, a fictitious mass ($\mu$) of 450 a.u. and timestep of 2.5 a.u. are used. The temperature is set at 400 K. These simulations were started from an ionic configuration from the end of a 11 ps of aqueous ferrous ion with U=0 eV, and we discarded the first 1 ps of the simulation. The constant of motion in Fig. 1 has negligible fluctuations as compared to the change in the potential energy. The lower left panel in Fig. 1 shows Fe-O radial distribution function for aqueous ferrous ion in the case with U=0 or 5.6 eV. Consistent with the static calculation in the previous Section, the radial distribution function with U=5.6 eV shows a first peak displaced to a larger distance. In the case of aqueous ferric ion, there is no significant change in the radial distribution function.

Moreover, we show in Fig. 2 the power spectra of Fe-O stretching vibrations between the ion and the first solvation shell oxygen atoms calculated from velocity-velocity correlation functions. With U=0 eV (upper panel of Fig. 2), the power spectra peak around 278 and 390 cm$^{-1}$, respectively, for ferrous (solid line) and ferric (dashed line) ions. Previous simulations with QM/MM methods [29] gave two stretching-mode frequencies 342 and 276 cm$^{-1}$ for ferrous ion, and 443 and 379 cm$^{-1}$ for ferric ion. Although the two stretching modes are not resolved in our calculations, our results are in good agreement with the previous study. In particular, we also observe an increase of about 100 cm$^{-1}$ in stretching frequencies in going from ferrous to ferric ion, which is due to the larger Coulomb interactions experienced by oxygen atoms surrounding the ferric ion. On the other hand, the Fe-O stretching frequencies for both ferrous and ferric ions are not affected by Hubbard U correction as shown in the lower panel of Fig. 2, despite the fact that the Fe-O distances increase in the presence of the Hubbard U correction for ferrous ion.

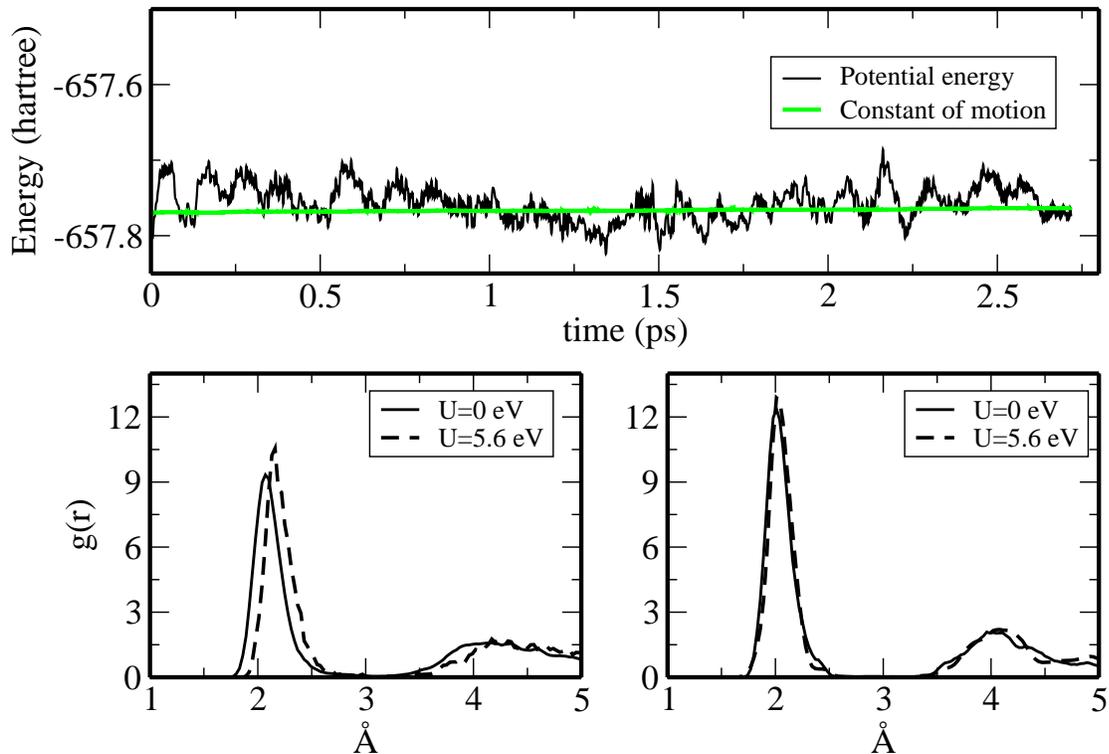

FIG. 1: Upper panel: The potential energy and the constant of motion of the simulation with ferrous ion solvated in 31 water molecules (U=5.6 eV). Lower left: the Fe-O radial distribution functions of aqueous ferrous ion with U=0 and 5.6 eV. Lower right: the Fe-O radial distribution functions of aqueous ferric ion with U=0 and 5.6 eV.

## VII. CONCLUSION

The DFT + Hubbard U approach to describe strong correlations in transition metal materials has been discussed in the context of Car-Parrinello molecular dynamics. The Hubbard correction has been shown to produce significant improvements on widely-used DFT functionals in describing transition metal oxides [10] and molecular complexes [15].

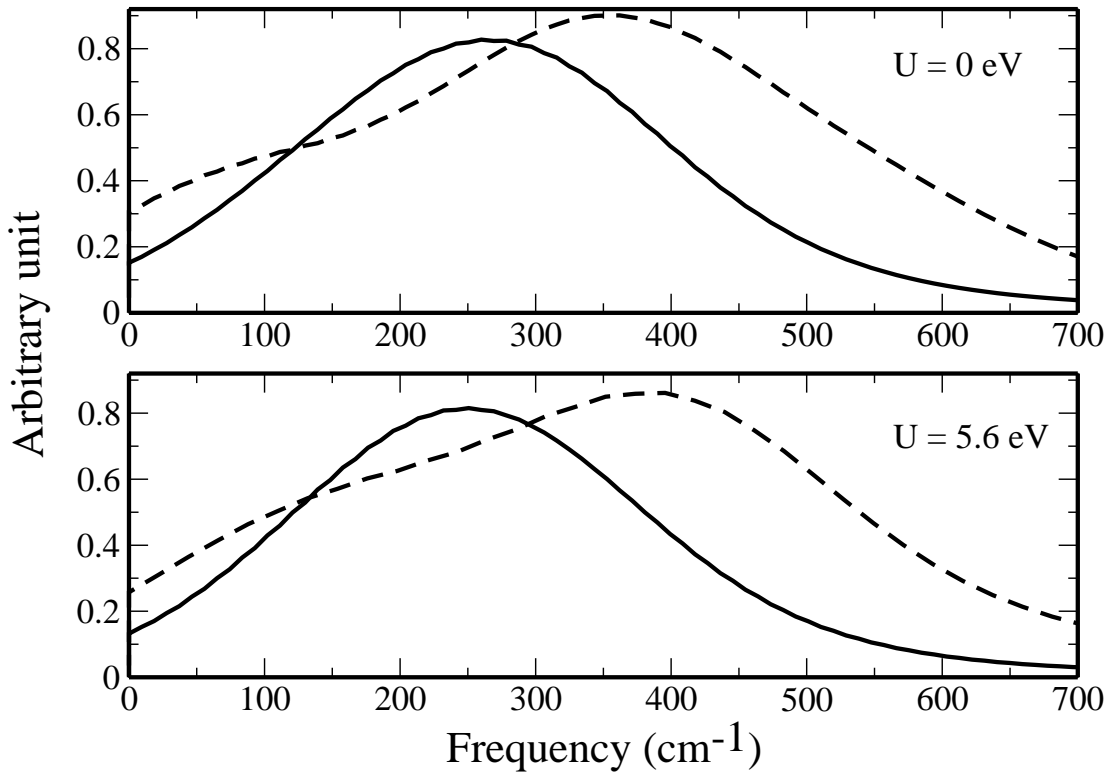

FIG. 2: Power spectra of ion-oxygen vibrational motions for ferrous (solid lines) and ferric (dashed lines) ions calculated from velocity-velocity correlation functions. Results from simulations with U=0 eV and U=5.6 eV are shown in the upper and lower panel, respectively.

Using linear-response calculations [14], we calculated the value of U for the hexa-aqua ferrous and ferric ions finding to be 5.6 eV in both cases. With this correction included, the inner-sphere reorganization energy is increased by 0.25 eV. This increase in reorganization energy is due to the change in equilibrium structure of the hexa-aqua ferrous ion alone. Moreover, we have implemented the Hubbard U correction in a Car-Parrinello molecular dynamics formalism, as implemented in QUANTUM-ESPRESSO [16]. The simulations for the aqueous ferrous and ferric ions show in the Fe-O radial distribution functions a similar elongation for the ferrous ion alone and bring our best estimate of the reorganization energy for this reaction to be 2.18 eV, in excellent agreement with the 2.1 eV experimental value [25]. Despite the changes in structure and energetics, the Fe-O stretching-mode frequencies are not affected by the Hubbard U correction.

We gratefully acknowledge support from the Croucher Foundation (P.H.-L.S.), MURI grant DAAD 19-03-1-0169 (N.M.) and the MRSEC Program of the National Science Foundation under the award number DMR 02-13282 (P.H.-L.S.). Computational facilities have been provided through NSF grant DMR-0414849 and PNNL grant EMSL-UP-9597.

---

[1] R. Car and M. Parrinello, Phys. Rev. Lett. **55**, 2471 (1985).

[2] K. Laasonen, A. Pasquarello, R. Car, C. Lee, and D. Vanderbilt, Phys. Rev. B **47**, 10142 (1993)

[3] D. Vanderbilt, Phys. Rev. B 41, 7892 (1990).

[4] J. P. Perdew, K. Burke, and M. Ernzerhof, Phys. Rev. Lett. **77**, 3865 (1996).

[5] M. Grüning, O. V. Gritsenko, S. J. A. van Gisbergen, and E. J. Baerends, J. Phys. Chem. A **105**, 9211 (2001).

[6] J. P. Perdew and A. Zunger, Phys. Rev. B **23**, 5048 (1981).

[7] V. I. Anisimov, editor. *Strong Coulomb Correlations in Electronic Structure Calculations*, Gordon and Breach, New York, 2000.

[8] V. I. Anisimov, J. Zaanen, and O. K. Andersen, Phys. Rev. B, **44**, 943 (1991).

[9] A. I. Liechtenstein, V. I. Anisimov, and J. Zaanen, Phys. Rev. B, **52**, R5467 (1995).

[10] M. Cococcioni, A. Dal Corso, and S. de Gironcoli, Phys. Rev. B, **67**, 094106 (2003).

[11] A. I. Liechtenstein, V. I. Anisimov, and J. Zaanen, Phys. Rev. B, **52**, R5467 (1995).

[12] V. I. Anisimov, F. Aryasetiawan, and A. I. Liechtenstein, J. Phys.: Condens. Matter, **9**, 767 (1997).

[13] W. E. Pickett, S. C. Erwin, and E. C. Ethridge, Phys. Rev. B, **58**, 1201 (1998).

[14] M. Cococcioni and S. de Gironcoli, Phys. Rev. B, **71**, 035105 (2005).

[15] H. J. Kulik, M. Cococcioni, D. Scherlis, and N. Marzari, Phys. Rev. Lett., **97**, 103001 (2006).

[16] S. Baroni, A. Dal Corso, S. de Gironcoli, P. Giannozzi, C. Cavazzoni, G. Ballabio, S. Scandolo, G. Chiarotti, P. Focher, A. Pasquarello, K. Laasonen, A. Trave, R. Car, N. Marzari, A. Kokalj, http://www.pwscf.org/.

[17] F. Zhou, C. A. Marianetti, M. Cococcioni, D. Morgan, and G. Ceder, Phys. Rev. B, **69**, 201101 (2004).

[18] F. Zhou, M. Cococcioni, C. A. Marianetti, D. Morgan, and G. Ceder, Phys. Rev. B, **70**, 235121 (2004).

[19] R. A. Marcus, J. Chem. Phys. **24**, 966 (1956).

[20] R. A. Marcus and N. Sutin, Biochem. Biophys. Acta **881**, 265 (1985).

[21] R. A. Kuharski, J. S. Bader, D. Chandler, M. Sprik, M. L. Klein, and R. W. Impey, J. Chem. Phys. **89**, 3248 (1988).

[22] P. H.-L. Sit, M. Cococcioni, and N. Marzari, Phys. Rev. Lett., **97**, 028303 (2006).

[23] G. J. Herman and G. W. Neilson, J. Phys.:Condens. Matter, **4**, 649 (1992).

[24] G. J. Herman and G. W. Neilson, J. Phys.:Condens. Matter, **4**, 627 (1992).

[25] K. M. Rosso and J. R. Rustad, J. Phys. Chem. A, **104**, 6718 (2000).

[26] G. Makov and M. C. Payne, Phys. Rev. B **51**, 4014 (1995).

[27] http://www.physics.rutgers.edu/ dhv/uspp/index.html.

[28] S. Amira, D. Spangberg, V. Zelin, M. Probst, and K. Hermansson, J. Phys. Chem. **109**, 14235 (2005).

[29] T. Remsungnen and B. M. Rode, Chem. Phys. Lett. **367**, 586 (2003).